\documentclass{ws-mpla}

\usepackage[hidelinks]{hyperref}

\begin{document}

\renewcommand{\draftnote}{}
\renewcommand{\trimmarks}{}

\markboth{E. A. Matute} {Low-scale minimal linear seesaw model for neutrino mass and flavor mixing}

\catchline{}{}{}{}{}

\title{Low-scale minimal linear seesaw model \\
for neutrino mass and flavor mixing}

\author{\footnotesize Ernesto A. Matute}

\address{Facultad de Ciencia, Departamento de F\'{\i}sica, Universidad de Santiago de Chile \\
(USACH), Santiago, Chile\\
\href{mailto:ernesto.matute@usach.cl}{ernesto.matute@usach.cl}}

\maketitle

\pub{}{}

\begin{abstract}
We consider an extension of the Standard Model with three right-handed (RH)
neutrinos and a Dirac pair of extra sterile neutrinos, odd under a discrete
$Z_2$ symmetry, in order to have left--right symmetry in the neutrino content
and obtain tiny neutrino masses from the latter ones only. Our working
hypothesis is that the heavy RH neutrinos do not influence phenomenology at
low energies. We use the usual high--scale seesaw to suppress all of the mass
terms involving RH neutrinos and a low-scale minimal variant of the linear
seesaw led by the Dirac mass of the extra sterile neutrinos to provide the
small mass of active neutrinos. One of the active neutrinos is massless,
which fixes the mass of the other two on the basis of a soft breaking of
the $Z_2$ symmetry. The mixing between the extra neutrinos makes for a
particle that effectively behaves like a Dirac sterile neutrino with mass
around the GeV level.

\end{abstract}

\ccode{PACS Nos.: 14.60.St, 14.60.Pq, 14.60.Lm, 13.15.+g}

\section{Introduction}
\label{Introduction}

The finding of neutrino masses and their flavor mixing together with
the existence of dark matter (DM) are major experimental facts that
ask for extensions of the Standard Model (SM). The addition of
right-handed (RH) neutrinos, singlets under the gauge symmetry of the
SM, seems to be the simplest input to bring about significant results.
They can be combined with the left-handed (LH) neutrinos to generate
Dirac mass terms ($m_D$) via Yukawa couplings with the Higgs doublet,
just like for charged leptons and quarks.

A minimalist approach is to introduce one RH neutrino per
generation. However, this requires extremely tiny Yukawa couplings to
explain the observed masses, which suggests the existence of a new
fundamental mass scale. This appears naturally in the high-scale
Type-I seesaw mechanism for neutrino mass
generation,\cite{Seesaw1}$^{\mbox{--}}$\cite{Seesaw5} where the RH
neutrinos are assumed to possess heavy Majorana masses ($m_R$).
In this scenario, the neutrino mass matrix in the basis
($\nu_L, \nu^c_R$) is given by
\begin{equation}
\mathcal{M}_\nu = \left(
\begin{array}{cc}
0 & m_D \\ m_D^T & m_R
\end{array} \right).
\label{massmatrixI}
\end{equation}
Here, $\nu^c_R= C \overline{\nu_R}^T$, using the notation of
Ref.~\refcite{notation}. The physical states after diagonalization of
the mass matrix include light ($\nu$) and heavy ($\nu_H$) sectors
of Majorana neutrinos with masses
\begin{equation}
m_\nu \simeq - m_D m^{-1}_R m^T_D , \qquad m_{\nu_H} \simeq m_R \gg m_\nu .
\label{Nemass1}
\end{equation}
From the point of view of model building, a natural expectation is a Dirac
neutrino mass $m_D$ similar in size to the Dirac mass of the charged lepton, i.e.
\begin{equation}
m_D \approx m_\tau \sim 1 \; \mbox{GeV}
\label{tauscale}
\end{equation}
for the third generation of neutrinos,
which leads to $m_\nu \sim 10^{-2}$ eV for $m_R \sim 10^{12}$ GeV.
In this scheme, the mixing between the active neutrinos and the
heavy RH neutrinos is of order $m_D / m_R \simeq \sqrt{m_\nu
/ m_{\nu_H}} \ll 1$, so that any experimental observation of flavor violation
involving charged leptons would imply the existence of new physics
at a much lower scale. Models with low Majorana masses for the RH neutrinos
were then proposed, in which the lightest Majorana RH neutrino plays
the role of a keV DM candidate.\cite{Shapo1,Shapo2}

Well-motivated variants of these schemes include the inverse
seesaw,\cite{Inverse1}$^{\mbox{--}}$\cite{Inverse3} the linear
seesaw,\cite{Linear1}$^{\mbox{--}}$\cite{Linear3} and the extended
seesaw,\cite{Extended1,Extended2} which are usually known as low-scale
seesaw mechanisms. These scenarios introduce an additional singlet
neutrino per generation, such that in the new basis ($\nu_L, \nu^c_R, S_L$)
the full mass matrix becomes
\begin{equation}
\mathcal{M}_\nu = \left(
\begin{array}{ccc}
0 & m_D & \mu^\prime_L \\ m^T_D & 0 & M_D \\
\mu^{\prime T}_L & M^T_D & \mu_L
\end{array} \right) ,
\label{massmatrixII}
\end{equation}
where $\mu^\prime_L$ and $M_D$ are the mass terms induced by
the couplings of the new fermion to the LH and RH neutrinos,
respectively, while $\mu_L$ is the new Majorana mass.

After block matrix diagonalization, the active neutrino mass has the
form
\begin{equation}
m_\nu \simeq m_D M^{T -1}_D \mu_L M^{-1}_D m^T_D
\label{Nemass2}
\end{equation}
in the inverse seesaw case, with $\mu^\prime_L=0$. Compared  to
Eq.~(\ref{Nemass1}), there is an extra suppression with respect to the
classical seesaw given by $m_\nu \simeq (m_D^2 / M_D)(\mu_L / M_D)$,
which leads to $M_D \sim 1 \; \mbox{TeV}$ for $m_\nu \sim 10^{-2}$ eV,
$m_D$ as in Eq.~(\ref{tauscale}), and $\mu_L \sim 10 \; \mbox{keV}$,
attributed to the smallness of lepton number violation.

In the linear seesaw mechanism, with $\mu_L = 0$, one has instead
\begin{equation}
m_\nu \simeq - m_D (\mu^\prime_L M^{-1}_D)^T - \mu^\prime_L M^{-1}_D m^T_D .
\label{Nemass3}
\end{equation}
Note here that there is an extra suppression with respect to the classical
seesaw according to $m_\nu \simeq (m_D^2 / M_D)(2 \mu_L^\prime / m_D)$. Now,
for a Dirac neutrino mass in accord with Eq.~(\ref{tauscale}), and
$\mu_L^\prime \sim 10 \; \mbox{keV}$ also attributed to the smallness of
lepton number violation, we obtain $m_\nu \sim 10^{-2}$ eV for
$M_D \sim 10^{3}$ TeV. The $M_D$ scale can be brought down to
the TeV range as done for the inverse seesaw, making the model more appealing,
by using the similarity condition with the first generation of charged
leptons instead of Eq.~(\ref{tauscale}), i.e. $m_D \approx m_e \sim 1 \;
\mbox{MeV}$, but maintaining $m_\nu \sim 10^{-2}$ eV and $\mu_L^\prime$ at
the keV scale.

Alternatively, in the extended seesaw scheme, with
$\mu^\prime_L=0$ and $\mu_L=0$, $m_\nu=0$ at the tree level. An extension of
all of these scenarios with new singlet neutrinos can lead to the appearance
of DM candidates.

No definitive evidence, however, has been observed of effects from
extensions of the SM with RH neutrinos. It is then natural to assume
that these particles are just elements of a large-scale physics,
preventing then their production at current high-energy accelerators.
Our working hypothesis in this letter is that the standard RH neutrinos
are superheavy, exactly as in the original classic seesaw, so that they
do not influence phenomenology at low energies and therefore they are
neither the relevant particles to generate the tiny neutrino mass of
active neutrinos nor the particles able to provide a DM candidate, in
contrast with the former seesaw schemes.

If neutrino masses have no relation to RH neutrinos, then
Eqs.~(\ref{Nemass1}), (\ref{Nemass2}), and (\ref{Nemass3}) should be
modified according to couplings of new sterile neutrinos incorporated to
the spectrum, suppressing the mass terms associated with RH neutrinos
($m_D$, $M_D$, and $m_R$), that is, a seesaw with sterile neutrinos
having a scale which is much lower than the scales of the above mechanisms,
so relaxing the restrictions of having to deal with heavy RH neutrinos to
explain the smallness of masses. Yet, as long as the new neutrinos are
simply singlets under the SM gauge group, we should end up with one of the
above-mentioned low-scale seesaw models.

Motivated by these expectations, we consider a new simple low-energy model
which gives explanations of the origin and nature of neutrino mass and
flavor mixing. We assume a seesaw mechanism where three RH neutrinos are
added to the SM with high-scale Majorana masses so that they become decoupled
generating highly suppressed terms for active neutrinos, just as in the
canonical high-scale seesaw mechanism. We think of RH neutrinos as part of a
large scale physics, featuring, for example, left--right symmetry, with very weak,
practically unobservable effects at low energies. Thus, the origin of the
smallness of neutrino masses would not be in the heavy RH neutrinos. It is an
alternative to seesaw mechanisms where neutrino masses depend on suppressed
Dirac mass terms which involve RH neutrinos, such as the canonical one and its
inverse and linear extended variants, among others published in the literature.

To realize a low-scale seesaw mechanism for the generation of neutrino mass,
without invoking the ordinary RH neutrinos, we simply introduce a Dirac pair
of sterile fermions, $N_R$ and $N_L$. Besides, in order to succeed in
completing this scheme, we consider a $Z_2$ symmetry under which the RH
neutrinos and all of the fermions in the SM have charge +1 (labeling the even
state), whereas the new sterile neutrinos have charge $-$1 (labeling the odd state).
Thus, while the usual low-scale seesaw mechanisms operate with singlet fermions
which are in the $Z_2$ even state, our model works with extra neutrinos in the
$Z_2$ odd state. A soft breaking of this symmetry would generate the
tiny mass of neutrinos.\cite{tHooft}

The three RH neutrinos, partners of the three LH neutrinos, are introduced to
restore left--right symmetry in the neutrino content of the $Z_2$ even sector.
The extra sterile neutrinos do not break this symmetry because they are in the
$Z_2$ odd state.

In order to understand the origin and soft breaking for the $Z_2$
symmetry considered here, we invoke the so-called presymmetry described in
Ref.~\refcite{Presym}. Presymmetry characterizes an underlying electroweak theory
of quarks and leptons based on the requirements of a left--right symmetry in
fermionic content and a charge symmetry between initially postulated fractional
electroweak charges. It is built upon the U$(1)_{B-L}$ global symmetry, which forbids
Majorana mass terms. Presymmetry is broken at the level of quarks and leptons,
making possible symmetry breaking terms such as the Majorana mass terms for RH
neutrinos and those related to the $Z_2$ symmetry. In a sense, presymmetry reflects
the high affinity of LH neutrinos with their RH partners before the breaking that
separates or dissociates them. It is the root of the $Z_2$ symmetry through which
we introduce the new leveling of singlet fermions. Presymmetry also explains why
the three RH neutrinos are required, despite them assumed to be so heavy that
they do not influence phenomenology at low energies.

We shall see that the extra  sterile neutrinos generate a massless neutrino within
a linear seesaw scheme, providing thus a simple solution to the neutrino mass and
mixing problems of the SM. To summarize, in this letter, we propose a low-scale
(at or below GeV) minimal variant of the linear seesaw mechanism in a scenario with
left--right symmetry in the neutrino content. We organize the work as follows.
In Sec.~\ref{SSRH}, we refer to the SM extended with RH neutrinos, which generate
neutrino masses via the usual high-scale seesaw mechanism. The extra sterile
neutrinos, which are protected by the discrete $Z_2$ symmetry, become passive
spectators. In Sec.~\ref{SSSN}, we examine the low-scale minimal variant of the
linear seesaw mechanism now involving the new sterile neutrinos, which, keeping
the heavy RH neutrinos decoupled and assuming a soft breaking of the $Z_2$
symmetry, induce the actual neutrino masses. Phenomenological aspects of the
model are discussed in Sec.~\ref{Pheno}. Conclusions are given in
Sec.~\ref{Conclusions}.

\section{Seesaw with RH Neutrinos Keeping Extra Sterile \\
Neutrinos Decoupled}
\label{SSRH}

We start with the SM extended with three RH neutrinos ($\nu_{R\alpha},
\alpha=e,\mu,\tau$) in the $Z_2$ even sector plus a Dirac pair of sterile
neutrinos ($N_R, N_L$) in the $Z_2$ odd sector, then having a left--right
symmetry in the neutrino content. The terms of the gauge-invariant Lagrangian
relevant to the neutrino masses are
\begin{equation}
- \mathcal{L} \supset {\it y}_\nu \overline{\ell_L} \,
\widetilde{\phi} \nu_R + \frac{m_R}{2} \overline{\nu^c_R} \nu_R
+ M_D \overline{N_L} N_R + h.c. ,
\label{YukawaRH}
\end{equation}
where, omitting flavor and isospin indexes, $\ell_L$ denotes the LH lepton
doublet and $\widetilde{\phi}=i \sigma_2 \phi^{*}$ represents the charge
conjugated field of the scalar Higgs doublet. We consider a Dirac mass term
for the extra sterile neutrinos and invoke the underlying presymmetry to forbid
their Majorana mass terms, allowing these only for RH neutrinos. Although they
can be generated via the vacuum expectation value of gauge singlet scalars that
couple to the gauge singlet fermions, in this paper we proceed with the effective
Lagrangian given in Eq.~(\ref{YukawaRH}). To make the point, we here neglect the
mixing of the extra sterile neutrinos with the LH and RH neutrinos. This Lagrangian
is invariant under the $Z_2$ symmetry and also consistent with conservation of
lepton parity $(-1)^L$,\cite{EMa} but, being established at the level
of leptons, it is not under presymmetry. Even though presymmetry breaking
allows the LH and RH neutrino couplings to change, Majorana mass terms for the
extra sterile neutrinos remain excluded.

After the electroweak symmetry breaking, Dirac mass terms with
$m_D = \it{y}_\nu \langle \phi^\circ \rangle$ are induced. In the
($\nu_L$, $\nu^c_R$, $N^c_R$, $N_L$) basis, the full 8x8 neutrino mass matrix
takes the form
\begin{equation}
\mathcal{M}_\nu = \left(
\begin{array}{cccc}
0 & m_D & 0 & 0 \\ m^T_D & m_R & 0 & 0 \\
0 & 0 & 0 & M_D \\ 0 & 0 & M_D & 0
\end{array} \right) \, .
\label{massmatrix}
\end{equation}

In the limit $m_R \gg m_D$, the diagonalization of the mass matrix
leads to light Majorana neutrinos ($\nu$) and a very heavy Majorana
neutrino ($\nu_H$), just as in the classical high-scale seesaw scenario,
together with a Dirac neutrino ($N$) that appears as an inactive
spectator. We have
\begin{equation}
m_\nu \simeq - m_D m^{-1}_R m^T_D , \qquad m_{\nu_H} \simeq m_R ,
\qquad M^\pm_N=\pm M_D ,
\end{equation}
where $m_\nu$ and $m_{\nu_H}$ are 3$\times$3 matrices.

The tiny neutrino mass would be the only trace at low energies left by
the decoupled heavy RH neutrinos. Moreover, an increase of the $m_R$ would
mean an increase of the $m_D$, breaking even more the expected similarity
of the Dirac mass terms of neutrinos and their charged leptonic weak
partners.

Extended models with additional singlet neutrinos in the $Z_2$ even state can
lower the scale of the $m_R$, as shown in Sec.~\ref{Introduction}. However,
this produces a proliferation of particles in a scenario where no signs of RH
neutrinos have been established yet. Other possibilities, where the $m_R$ are
naturally held up at a high scale, are then worth considering, giving up the
idea that the small neutrino mass has to do with the suppressed Dirac mass
term $m_D$, as in the seesaw mechanisms we have referred to previously.

\section{Seesaw with Extra Sterile Neutrinos Having RH \\
Neutrinos Decoupled}
\label{SSSN}

We now consider the low-scale seesaw mechanism for neutrino mass generation
led by the extra sterile neutrinos. New terms are introduced in the Lagrangian
switching on the tiny mixing of these neutrinos with the active neutrinos and
the RH neutrinos. The $Z_2$ symmetry that differentiates the new sterile
neutrinos from the others is broken softly. The reading of the Lagrangian
in Eq.~(\ref{YukawaRH}) extended with the new mixing terms is
\begin{eqnarray}
- \mathcal{L} &\supset& {\it y}_\nu \overline{\ell_L} \,
\widetilde{\phi} \nu_R + {\it y}^\prime_\nu \overline{\ell_L} \,
\widetilde{\phi} N_R + {\it y}^\prime_L \overline{\ell_L} \,
\widetilde{\phi} N^c_L + \frac{m_R}{2} \overline{\nu^c_R} \nu_R
\nonumber \\ && + M_D \overline{N_L} N_R
+ \mu^\prime_D \overline{N_L} \nu_R
+ \mu^\prime_R \overline{N^c_R} \nu_R
+ h.c.
\label{YukawaDM}
\end{eqnarray}
As explained above, all of the terms in this Lagrangian that produce soft
breaking of the $Z_2$ and U$(1)_{B-L}$ symmetries are consistent with
the breaking of presymmetry.

When the Higgs field gets its vacuum expectation value, the mass
matrix that follows from Eq.~(\ref{YukawaDM}) in the
($\nu_L$, $\nu^c_R$, $N^c_R$, $N_L$) basis can be written in blocks as
\begin{equation}
\mathcal{M}_\nu = \left(
\begin{array}{cccc}
0 & m_D & m^\prime_D & \mu^\prime_L \\ m^T_D & m_R & \mu^\prime_R
& \mu^\prime_D \\ m^{\prime T}_D & \mu^{\prime T}_R & 0 & M_D \\
\mu^{\prime T}_L & \mu^{\prime T}_D & M_D & 0
\end{array} \right) ,
\label{massmatrixDM}
\end{equation}
where $m_D = \it{y}_\nu \langle \phi^\circ \rangle$,
$m^\prime_D = \it{y}^\prime_\nu \langle \phi^\circ \rangle$,
and $\mu^\prime_L = \it{y}^\prime_L \langle \phi^\circ \rangle$.
Note that $m_D$ and $m_R$ are 3$\times$3 matrices, while $m^\prime_D$ and
$\mu^\prime_L$ together with $\mu^\prime_R$ and $\mu^\prime_D$ are
3$\times$1 matrices. Comparing Eqs.~(\ref{massmatrix}) and
(\ref{massmatrixDM}), we see that the smallness of the added
masses does arise from the $Z_2$ symmetry, which becomes a good
symmetry as the new parameters go to zero. We use the high-scale
seesaw mechanism under the assumption that the Majorana masses
are larger than all of the other mass parameters.

The block diagonalization of $\mathcal{M}_\nu$ leads to a mass
matrix, which after neglecting all the suppressed seesaw contributions
(e.g. $m_D^2/m_R \ll m_D^\prime, \mu_L^\prime$), is reduced to the
following form in the ($\nu_L$, $N^c_R$, $N_L$) basis
\begin{equation}
\mathcal{M}^\prime_\nu = \left(
\begin{array}{ccc}
0 & m^\prime_D & \mu^\prime_L \\ m^{\prime T}_D & 0 & M_D \\
\mu^{\prime T}_L & M_D & 0
\end{array} \right) .
\label{effmassmatrix}
\end{equation}
This is the mass matrix left when the line and the column of $m_R$
in the mass matrix in Eq.~(\ref{massmatrixDM}) are removed, which
should be contrasted with that in Eq.~(\ref{massmatrixII}) relative
to the linear seesaw. It is equivalent to having in the neutrino
sector the effective Lagrangian
\begin{equation}
- \mathcal{L}_{eff} \supset {\it y}^\prime_\nu \overline{\nu_L} \,
\phi^\circ N_R + {\it y}^\prime_L \overline{\nu_L} \, \phi^\circ
N^c_L + M_D \overline{N_L} N_R + h.c. \, ,
\label{effYukawa}
\end{equation}
i.e. the Yukawa sector of the SM augmented with terms involving the extra
sterile neutrinos. Thus, no effects of RH neutrinos in the $Z_2$ even
state remain at low energies, in agreement with the null experimental
evidence so far proving the existence of RH neutrinos. The mass
hierarchy is now $m^\prime_D , \mu^\prime_L \ll M_D$, with
$m^\prime_D \sim \mu^\prime_L$.

The block diagonalization of $\mathcal{M}^\prime_\nu$ leads to the 3$\times$3
light neutrino mass matrix
\begin{equation}
m_\nu \simeq - \frac{m^\prime_D \mu^{\prime T}_L}{M_D}
- \frac{\mu^\prime_L m^{\prime T}_D}{M_D}
\label{light}
\end{equation}
and the heavier neutrino masses
\begin{eqnarray}
M^+_N &\simeq& M_D + \frac{(m^\prime_D + \mu^\prime_L)^T
(m^\prime_D + \mu^\prime_L)}{2 M_D} , \nonumber \\ && \label{heavy} \\
M^-_N &\simeq& - M_D - \frac{(m^\prime_D - \mu^\prime_L)^T
(m^\prime_D - \mu^\prime_L)}{2 M_D} . \nonumber
\end{eqnarray}
Equation~(\ref{light}) implies that just one of the three active neutrinos is
massless at the tree level, without constraining the values of the active--sterile
mixing couplings, as it can be seen if the determinant of the matrix $m_\nu$
is calculated in general (it is nonzero in the case of two generations).
Quantum corrections to this null mass eigenvalue turn out to be vanishingly
small.\cite{minimalseesaw} Note that as expected from the null trace of the
mass matrix, $\mathcal{M}^\prime_\nu$, $tr(m_\nu) + M^+_N + M^-_N =0$. Also, the
mass splitting of the pseudo-Dirac fermion goes with the neutrino masses. In
other words, the neutrino masses remove the degeneracy of the pair composing
what otherwise would be a Dirac particle. As already mentioned, the smallness
of $m^\prime_D$ and $\mu^\prime_L$, with $m^\prime_D \sim \mu^\prime_L$, is
protected by the $Z_2$ symmetry, and when they are small, $M_D$ is small too,
lowering the scale of the seesaw mechanism. The flavor states will be
combinations of the three SM-like neutrino mass eigenstates ($\nu$) and the
heavier mostly sterile pseudo-Dirac neutrino~($N$).

The resulting mass matrices in Eqs.~(\ref{effmassmatrix}) and (\ref{light}), as
well as the eigenvalues in Eq.~(\ref{heavy}), are very similar in form to the ones
found in the standard linear seesaw mechanism, with the difference that the texture
of our matrix is consequence of a different set up, which ends up with mass terms at
a much lower scale, without including the Dirac mass of RH neutrinos. Here, it is
important to remark the differences with such a linear seesaw given the close resemblance
between Eqs.~(\ref{Nemass3}) and (\ref{light}). One might argue that the whole procedure
of adding extra sterile neutrinos and decoupling the RH neutrinos is a mere relabeling
of the linear seesaw (albeit with additional decoupled particles), i.e. $\nu_R
\rightarrow N_R$, then obtaining the neutrino mass matrix simply by replacing the
notation $m_D \rightarrow m^\prime_D$.

However, it is not just a matter of labeling differently. In fact,
the usual linear seesaw operates with the singlet fermions $\nu_R, S_L$, which are in
the $Z_2$ even state, while our low-scale minimal linear seesaw model works with the extra
neutrinos $N_R, N_L$, which are in the $Z_2$ odd state. Thus, when we compare the mass
terms $m_D \overline{\nu_L} \nu_R$ (in the standard linear seesaw) and $m_D^\prime
\overline{\nu_L} N_R$ (in our model) to each other, for instance, we cannot claim that
all of this is just a change of notation: there is invariance under $Z_2$ symmetry
transformations ($\nu_{L,R} \rightarrow \nu_{L,R}; N_{R,L} \rightarrow - N_{R,L}$) in
the former case, but not in the latter one. Moreover, under a soft breaking of $Z_2$
we have $m_D^\prime \ll m_D$. Clearly, the two models we are referring to are not the same.

Furthermore, the seesaw mass matrix given in Eq.~(\ref{effmassmatrix}) contains
elements ($m_D^\prime, \mu_L^\prime$) associated with the sterile neutrinos $N_R, N_L$
in the $Z_2$ odd state, while the usual linear seesaw mass matrix contains those
($m_D, \mu_L^\prime$) related to the singlet fermions $\nu_R, S_L$ in the $Z_2$ even state.
Being $m^\prime_D \sim \mu^\prime_L \ll m_D$, the mass of the heavy Dirac neutrino ($M_D$)
in the new variant is much lower than the mass of the heavy neutrino predicted by the
standard linear seesaw.

Note in particular that in the known linear seesaw, the linearity is because of $m_D$,
which in our context is $m_D^\prime$. Besides, the usual linear seesaw does introduce
three pairs of additional neutrinos, in contrast to the realization of the low-scale minimal
seesaw presented in this paper, where a single pair of extra neutrinos is sufficient to
explain the light neutrino masses. And even though this last statement is well established
in minimal seesaw models,\cite{MiniSS1}$^{\mbox{--}}$\cite{MiniSS3} also with singlet neutrinos
that can have masses below the TeV scale, the light neutrino masses still depend on the Dirac
mass term $m_D$ with RH neutrinos in the $Z_2$ even state, which in our case goes away with the
decoupling of RH neutrinos via the high-scale seesaw mechanism.

On the other hand, now regarding the neutrino mass in Eq.~(\ref{light}) from the
phenomenological point of view, there is an additional suppression with respect to the
classical seesaw and the usual linear seesaw (see Eqs.~(\ref{Nemass1}) and (\ref{Nemass3}))
given by
\begin{equation}
m_\nu \simeq \frac{m_D^2}{M_D} \cdot \frac{2 \mu_L^\prime}{m_D} \cdot \frac{m_D^\prime}{m_D} ,
\end{equation}
which leads to $M_D \sim 10 \; \mbox{GeV}$ for $m_\nu \sim 10^{-2} \; \mbox{eV}$ and
$\mu_L^\prime \sim m_D^\prime \sim 10 \; \mbox{keV}$, now attributed to the soft breaking of
lepton number conservation and $Z_2$ symmetry. This means $m_D^\prime \sim 10^{-5} \, m_D$
for $m_D$ in accord with the similarity condition in Eq.~(\ref{tauscale}), and thus a seesaw
scale ($M_D$) which is about 5 orders of magnitude smaller than the scale of the usual linear
seesaw. This can be shrunk to $m_D^\prime \sim 10^{-2} \, m_D$ if
Eq.~(\ref{tauscale}) is replaced by the similarity condition involving the first generation
of charged leptons. But we keep the $Z_2$ symmetry breaking terms at the keV scale, same as
done for the inverse and linear seesaw mechanisms, making the model more appealing and stressing
its differences. As argued in Sec.~\ref{Introduction}, the soft breaking of the $Z_2$ symmetry
would have its origin in the presymmetry breaking, allowing then the coupling of LH and RH
neutrinos to the extra sterile neutrinos. Interesting enough, the new variant has a seesaw scale
at the GeV range independently of the similarity condition used between the size of the Dirac
neutrino mass ($m_D$) and the Dirac mass of the charged lepton, making thus a difference with the
others. Whatever view is taken on this, we can state unequivocally that we here have a low-scale
(at or below the GeV) minimal variant of the linear seesaw mechanism within a scenario with
left--right symmetry in the neutrino content that has not been dealt with yet.

\section{Phenomenological Aspects of the Model}
\label{Pheno}

The neutrino phenomenology at low energies relies drastically upon the extra sterile
neutrinos in the $Z_2$ odd state, whose couplings to the active neutrinos in the
even state determine their decay channels and lifetimes.

The prediction that the lightest active neutrino is massless excludes the
possibility of having a quasi-degenerate mass spectrum and fixes the masses of
the other two active neutrinos to be (8.6$\times$10$^{-3}$, 0.05) eV in the normal hierarchy
($m_1=0$, $m_2<m_3$) or (4.92$\times$10$^{-2}$, 0.05) eV in the inverted hierarchy
($m_3=0$, $m_1<m_2$),\cite{PDG} almost doubling the total sum of neutrino masses
of the normal ordering.

Based on these results, we now revise the parameter space of our model starting at
the keV scale to see if the heavy sterile neutrino in the $Z_2$ odd state classifies
as a DM candidate. Its decay into the lighter SM particles in the even state is a
fact that has to be pondered as the $Z_2$ symmetry is broken softly. We need to make
certain that its lifetime is longer than the age of the Universe. We also have to
corroborate that this heavy neutrino that drives the seesaw complies with the
constraint on DM relic abundance. Since the mass scale of $N$ is much higher than the
mass splitting, the Dirac pair of sterile neutrinos becomes approximately degenerate
and it can be treated as a Dirac neutrino.

The dominant decay channel of the heavy neutrino would be
$N \rightarrow 3 \nu$ through active--sterile neutrino mixing and weak interaction
of $\nu$. Neglecting the tiny difference between the two Majorana sterile neutrinos
and so the neutrino masses, the decay width of this decay mode is given
by\cite{decay1,decay2}
\begin{equation}
\Gamma_{N\rightarrow 3\nu} \simeq \frac{G_F^2 M_N^5}{96 \pi^3} |\Theta_{\nu N}|^2 ,
\end{equation}
where $|\Theta_{\nu N}|$ is the small mixing parameter between the active and sterile
neutrinos defined in our model as
\begin{equation}
|\Theta_{\nu N}|^2 = \sum_{\alpha=e,\mu,\tau} \left( |\Theta_{\nu_\alpha N_L}|^2 +
|\Theta_{\nu_\alpha N_R}|^2 \right) ,
\label{Theta}
\end{equation}
in which
\begin{equation}
\Theta_{\nu_{\alpha}N_L} \simeq \frac{m^\prime_{D\alpha}}{M_D} , \qquad
\Theta_{\nu_{\alpha}N_R} \simeq \frac{\mu^\prime_{L\alpha}}{M_D} .
\label{mixings}
\end{equation}
With $M_N$ in the 10 keV range, it leads to
\begin{equation}
\Gamma_{N\rightarrow 3\nu} \simeq \frac{|\Theta_{\nu N}|^2}{1.4 \times 10^{14} \; \mbox{s}}
\left( \frac{M_N}{10 \; \mbox{keV}} \right)^5 .
\end{equation}
Thus, from the lifetime of $N$ defined as $\tau_N = 1 / \Gamma_N$, and an age for the
Universe of about $4.4 \times 10^{17}$ s,\cite{ageU} we obtain the
following bound on the active--sterile mixing:
\begin{equation}
|\Theta_{\nu N}|^2 \ll 3.3 \times 10^{-4} \left( \frac{10 \; \mbox{keV}}{M_N} \right)^5 .
\label{DMbound}
\end{equation}

An additional well-known decay channel is the radiative decay
$N \rightarrow \nu \gamma$, with a width given by\cite{decay2}
\begin{equation}
\Gamma_{N\rightarrow \nu \gamma} \simeq \frac{27 \alpha}{8\pi} \;
\Gamma_{N\rightarrow 3\nu} = \frac{1}{128} \; \Gamma_{N\rightarrow 3\nu} ,
\end{equation}
which predicts an outgoing photon energy of around half the mass of the
sterile neutrino, thus suggesting a keV line in the spectra of galaxies
to be searched by X-ray telescopes.\cite{Xray}

In the early Universe, the production of our sterile neutrinos would have
been dominated by the so-called non-resonant Dodelson--Widrow
mechanism,\cite{DWmechanism} which occurs at high temperatures due to the
active--sterile neutrino mixing. Since the sterile neutrinos do not
experience the SM forces and their mixing angles are so small, they would
not have been in thermal equilibrium with the known particles and produced
then in non-equilibrium processes. The abundance of DM is fixed by the DM
density $\Omega_{\mbox{\scriptsize DM}} h^2 = 0.12$,\cite{ageU} where $h$ is the reduced Hubble
constant. Assuming that the sterile neutrino is the sole DM particle, it is
given by\cite{relicDM}
\begin{equation}
\Omega_{\mbox{\scriptsize DM}} h^2 \sim 0.12 \left( \frac{|\Theta_{\nu N}|^2}{10^{-9}} \right)
\left( \frac{M_N}{10 \; \mbox{keV}} \right)^2 .
\label{DMdensity}
\end{equation}

On the other hand, we can make (see Ref.~\refcite{Shapo2})
\begin{equation}
|\Theta_{\nu N}|^2 = \frac{1}{M_D^2} \sum_{\alpha=e,\mu,\tau} \left(
|m^\prime_{D\alpha}|^2 + |\mu^\prime_{L\alpha}|^2 \right)
= \frac{1}{M_N} tr(m_\nu) = \frac{1}{M_N} \sum_{i=1}^{3} m_i .
\label{mixconstraint}
\end{equation}
If we choose $M_N \simeq 7$ keV as a possible DM sterile neutrino mass and
adopt the normal hierarchy for the active neutrino masses, we get
$|\Theta_{\nu N}|^2 \simeq 0.7 \times 10^{-5}$ and so $|m^\prime_{D\alpha}|$,
$|\mu^\prime_{L\alpha}|$ of a few eV, which complies with the bound in
Eq.~(\ref{DMbound}) but it is too high to do it with the DM condition. Thus,
the possibility that the Dirac sterile neutrino in the $Z_2$ odd state be a DM
candidate at the keV scale is ruled out, as expected.\cite{Shapo2} The sterile
neutrinos cannot be at the eV range either because of the cosmological bound on
the number of relativistic neutrino species.\cite{PDG}

We therefore assume $|m^\prime_{D\alpha}|$ and $|\mu^\prime_{L\alpha}|$ at the
keV level as in the usual low-scale inverse and linear seesaw schemes. For
illustrative purposes, we establish benchmark points choosing
$|m^\prime_D| \sim |\mu^\prime_L| \sim 4$ keV and the normal hierarchy
for neutrino masses. This leads to
\begin{equation}
M_N = \frac{\sum_{\alpha} \left( |m^\prime_{D\alpha}|^2 +
|\mu^\prime_{L\alpha}|^2 \right)}{\sum_{i} m_i} \sim 1 \; \mbox{GeV} ,
\end{equation}
that is, a sterile neutrino at the GeV scale, consistent with the value obtained
in Sec.~\ref{SSSN}. In contrast, the inverse and linear seesaws expect a rich
phenomenology just at or above the TeV level.

Note in particular that from Eq.~(\ref{light}) we have $m_D^\prime / M_D \sim
m_\nu / \mu^\prime_L$ which, when compared with $m_D / M_D \sim
m_\nu / \mu^\prime_L$ and $(m_D / M_D)^2 \sim m_\nu / \mu_L$ obtained
from Eqs.~(\ref{Nemass3}) and (\ref{Nemass2}), respectively, indicates a much
lower scale for the seesaw mechanism ($M_D$) if $\mu^\prime_L$
and $\mu_L$ are taken at the keV range, with
$m_D^\prime \sim \mu^\prime_L \ll m_D$.

It is also worth mentioning that the almost Dirac nature of the pair of heavy
neutrinos means no significant impact on the lepton number violating processes
such as the neutrinoless double beta decays. It can, however, mediate lepton flavor
violating processes like the $\mu \rightarrow e \gamma$ transition. But, the
active--sterile neutrino mixing angles are not large enough to be at the range of
the current experimental sensitivity.\cite{TeVne} Yet the GeV sterile neutrino
is potentially accessible directly by collider experiments.\cite{GeVne}

Finally, we should stress that the superheavy RH neutrinos are decoupled from
active neutrinos and therefore do not contribute to the low seesaw matrix in
Eq.~(\ref{light}). Nevertheless, they can give rise to the baryon asymmetry of the
Universe through high-scale leptogenesis.\cite{leptogenesis} We here note that a
low-scale resonant leptogenesis with the two extra sterile neutrinos, nearly
degenerate in their masses, is also viable.\cite{minimalseesaw}

\section{Conclusions}
\label{Conclusions}

We have extended the SM with three RH neutrinos, $\nu_R$, plus a Dirac pair
of sterile neutrinos, $N_R,N_L$, as a way to have left--right symmetry in neutrino
content and produce small neutrino masses just from the last ones. A discrete
$Z_2$ symmetry was considered, under which all particles have charge +1, except
the extra sterile neutrinos which have charge $-$1. It is a minimal extension in the
sense that additional Dirac pairs of sterile neutrinos in the odd state can be
included. We have argued that this $Z_2$ symmetry would have its origin
in the so-called presymmetry,  which typifies an underlying electroweak theory of
quarks and leptons with left--right symmetry in fermionic content and initially
postulated symmetric fractional charges, broken at the level of quarks and leptons.
This presymmetry, built on the global U$(1)_{B-L}$ symmetry, also guarantees that
the extra sterile neutrinos cannot have Majorana mass terms.

We have used a seesaw with superheavy RH neutrinos without raising the
Dirac masses connected with electroweak symmetry breaking, avoiding thus to alter
the natural similarity between neutrinos and charged leptons. This similitude has
been maintained and the RH neutrino mass terms suppressed via the standard
high-scale seesaw, so that the contribution of RH neutrinos becomes too small to be
worth consideration. In contrast to most of known type-I seesaw mechanisms, our
working hypothesis is that the RH neutrinos in the $Z_2$ even state do not influence
phenomenology at low energies. This effectively reduces the content of the seesaw
mechanism to three active neutrinos in the even state plus two sterile neutrinos
in the odd state. Yet, the three RH neutrinos are required as presymmetry
is demanded.

The main prediction of the minimal seesaw is that one of the active neutrinos is
massless, which fixes the small mass of the other two; in a more indicative sense,
three fermion generations are required to generate a massless neutrino. It has been
shown that this smallness of neutrino masses can be produced through a low-scale
seesaw with the Dirac pair of sterile neutrinos in the odd state. This couple
becomes a pseudo-Dirac fermion with a mass splitting given by the sum of light
neutrino masses. To explain the weakness of its coupling to the active neutrinos,
we have introduced a soft breaking for the $Z_2$ symmetry. We have
explained that this would have its origin in the presymmetry breaking, which allows
the coupling of LH and RH neutrinos to change, but keeping forbidden the Majorana
mass terms for the extra sterile neutrinos.

We have made clear that the almost Dirac sterile neutrino leading to the tiny neutrino
masses cannot be at the keV scale because it does not give the correct DM relic abundance.
In addition, as a low-scale alternative with no RH neutrinos to the standard linear and
inverse seesaw mechanisms, the scale of the sterile neutrino should lie well below the TeV
scale, that is, about the GeV range if we take into account that the small $Z_2$ symmetry
breaking terms should be at the keV scale, same as done for them, making the
model more appealing and stressing its differences. As a matter of fact, this new variant
has the seesaw scale at the GeV level independently of the similarity condition used
between the Dirac neutrino mass and the Dirac mass of the charged lepton, making therefore
a difference with those ones.

The resulting light neutrino mass matrix is very similar in form to the mass matrix
of the linear seesaw mechanism, albeit with a single pair of sterile neutrinos instead
of three and a different parameter regime, implying the lightest active neutrino to be
massless at tree level. Because of the similarity, one can assert that the seesaw under
consideration is a low-scale minimal variant of the linear seesaw. It is worth emphasizing,
however, that they are not the same. The usual linear seesaw operates with singlet fermions
which are in the $Z_2$ even state, while our low-scale minimal linear seesaw model works
with the extra neutrinos which are in the $Z_2$ odd state. Moreover, the mass of the heavy
Dirac neutrino, $M_D$, in the new variant turns out to be much lower than the mass of the
heavy neutrino predicted by the usual linear seesaw. Phenomenological aspects of the minimal
approach, related to lepton number violation and lepton flavor violation processes, have
been discussed in the literature.\cite{TeVne}

On the other hand, although the superheavy RH neutrinos are irrelevant to the generation
of neutrino masses, they can give rise to the baryon asymmetry of the Universe through
high-scale leptogenesis. A low-scale resonant leptogenesis with the two extra sterile
neutrinos is also viable as they are nearly degenerate in their masses.

Experimental tests of the model would be the finding of
active neutrino masses at the predicted scale and the appearance of a Dirac sterile
neutrino at the GeV scale in the particle spectrum. Our minimal seesaw also implies no
phenomenology associated with ordinary RH neutrinos with mass in the keV to TeV range,
as proposed by the alternative low-scale seesaw models that require RH neutrinos.
Here, we remark that these RH neutrinos are the ones that would be charged under a
U$(1)_{B-L}$ gauge symmetry and an eventual SU$(2)_L \times $SU$(2)_R \times $U$(1)_{B-L}$
gauge extension of the SM,\cite{LRsym1}$^{\mbox{--}}$\cite{LRsym3} in contrast to our
sterile neutrinos which would remain singlets in such a scenario. The whole scheme
proposed in this work would be ruled out if signals of RH weak currents are found at
low energies.

The puzzle of the observed mixing between leptons was not addressed by this paper.
The simple structure of the mass matrix that comes from the model, however, allows
to accommodate straightforwardly the so-called tri-bimaximal mixing pattern\cite{TBM}
as well as its deviations according to experimental data without invoking non-abelian
discrete flavor symmetries and symmetry-breaking scalar flavon fields, in the first
place. Analysis of these possibilities will be presented elsewhere.

\section*{Acknowledgments}

This work was partially supported by Vicerrector\'{\i}a de Investigaci\'on,
Desarrollo e Innovaci\'on, Universidad de Santiago de Chile (USACH).


\end{document}